\documentclass [11pt] {article}
\begin{document}
\begin{titlepage}
\setcounter{page}{1}

% define title
\begin{large}
\title{Dominance of the Nucleon Decay K-channels in $P_{LR}$ invariant
  $SO(10)$}
\author{Yoav Achiman and Susanne Bielefeld\\[0.5cm]
        Department of Physics\\
        University of Wuppertal\\
        Gau\ss{}str.~20, D--42097 Wuppertal \\
        Germany\\[1.5cm]}

\date{May 1997}

\maketitle
\setlength{\unitlength}{1cm}
\begin{picture}(5,1)(-12.5,-12)
\put(0,0){WUB 97-19}
\end{picture}
\parindent0cm

\begin{abstract}
RH rotations of the fermionic mass matrices are not observable in the
standard model, but dictate the details of the proton decay in GUTs. 
$P_{LR}$ symmetry in the two light families leads to a full RH mixing.
We give two versions of suitable broken $P_{LR}$ invariant $SO(10)$
which leads to nucleon decay modes similar to those of SUSY-GUTs with
rates in the range of observability of superKamiokande and ICARUS.
\end{abstract}
\thispagestyle{empty}
\end{large}

\end{titlepage}
\clearpage
\setcounter{page}{1}
%\end{document}
%

%\documentclass [11pt]{article}

%\textwidth=15.0 true cm
%textheight=23.0 true cm
%\voffset=1.5cm
%\hoffset=-0.5cm
%\oddsidemargin=1cm
%\evensidemargin=1cm

%\parindent 0pt
%\parskip 14pt
%\hfuzz 1cm
%\begin{document}
\newcommand{\bye}{\end{document}}
\newcommand{\ns}{\vfill\eject}
\newcommand{\be}{\begin{equation}}
\newcommand{\ee}{\end{equation}}
\newcommand{\beq}{\begin{eqnarray}}
\newcommand{\eeq}{\end{eqnarray}}
\newcommand{\plr}{$P_{LR}$}
\newcommand{\pd}{proton decay}
\newcommand{\llra}{\longleftrightarrow}
\newcommand{\ra}{\rightarrow}
\newcommand{\lra}{\longrightarrow}
\newcommand{\ov}{\overline}
\newcommand{\muk}{\mu^+ K^0}
\newcommand{\epi}{e^+\pi^0}
\newcommand{\et}{e^+\eta}
\newcommand{\ek}{e^+K^0}
\newcommand{\mpi}{\mu^+\pi^0}
\newcommand{\mt}{\mu^+\eta}
\newcommand{\ero}{e^+\rho^0}
\newcommand{\eo}{e^+\omega}
\newcommand{\eks}{e^+K^{*0}}
\newcommand{\mro}{\mu^+\rho^0}
\newcommand{\mo}{\mu^+\omega}
\newcommand{\nep}{\bar\nu_e\pi^+}
\newcommand{\nek}{\bar\nu_e K^+}
\newcommand{\nmpi}{\bar\nu_\mu\pi^+}
\newcommand{\nmk}{{\bar\nu}_\mu K^+}
\newcommand{\nero}{\bar\nu_e\rho^+}
\newcommand{\neks}{\bar\nu_e K^{*+}}
\newcommand{\nmro}{\bar\nu_\mu\rho^+}
\newcommand{\nmks}{\bar\nu_\mu K^{*+}}
%----------------------------------
\newcommand{\epin}{e^+\pi^-}
\newcommand{\mpin}{\mu^+\pi^-}
\newcommand{\eron}{e^+\rho^-}
\newcommand{\mron}{\mu^+\rho^-}
\newcommand{\nepn}{\bar\nu_e\pi^0}
\newcommand{\nekn}{\bar\nu_e K^0}
\newcommand{\nmpin}{\bar\nu_\mu\pi^0}
\newcommand{\nmkn}{{\bar\nu}_\mu K^0}
\newcommand{\neron}{\bar\nu_e\rho^0}
\newcommand{\neksn}{\bar\nu_e K^{*0}}
\newcommand{\nmron}{\bar\nu_\mu\rho^0}
\newcommand{\nmksn}{\bar\nu_\mu K^{*0}}
\newcommand{\neon}{\bar\nu_e \omega}
\newcommand{\nmon}{\bar\nu_\mu \omega}
\newcommand{\netn}{\bar\nu_e \eta}
\newcommand{\nmtn}{\bar\nu_\mu \eta}
\newcommand{\ci}[1]{~\cite{#1}} 
\def\NPB#1#2#3{Nucl. Phys. {\bf B} {\bf#1} (19#2) #3}
\def\PLB#1#2#3{Phys. Lett. {\bf B} {\bf#1} (19#2) #3}
\def\PRD#1#2#3{Phys. Rev. {\bf D} {\bf#1} (19#2) #3}
\def\PRL#1#2#3{Phys. Rev. Lett. {\bf#1} (19#2) #3}
\def\PRT#1#2#3{Phys. Rep. {\bf#1} C (19#2) #3}
\def\ARAA#1#2#3{Ann. Rev. Astron. Astrophys. {\bf#1} (19#2) #3}
\def\ARNP#1#2#3{Ann. Rev. Nucl. Part. Sci. {\bf#1} (19#2) #3}
\def\MODA#1#2#3{Mod. Phys. Lett. {\bf A} {\bf#1} (19#2) #3}
\def\NC#1#2#3{Nuovo Cim. {\bf#1} (19#2) #3} 
\def\ANPH#1#2#3{Ann. Phys. {\bf#1} (19#2) #3} 

\parindent0cm 

New results are expected in the near future from the proton decay
experiments superKamiokande~\cite{sk} and ICARUS~\cite{IC}. This is the reason for a
new wave of papers~\cite{pd1}, \cite{pd2} discussing proton decay, mainly in
supersymmetric (SUSY) theories. \\ 

The  decay  of the proton was first predicted in the framework of
Pati-Salam $SU_C(4) \times SU_L(2) \times SU_R(2)$ L-R symmetric gauge
group~\cite{ps}  and $SU(5)$  GUT. The minimal $SU(5)$ which requires
unification of the gauge coupling constants around $M_{SU(5)} \approx
10^{14}\, GeV$  predicts gauge boson mediated proton decay, dominantly into
$e^+ {\pi}^0$ with a rate of
$$
\Gamma_{SU(5)} (P \to \epi)  \simeq 10^{-31 \pm 1}  \quad yrs^{-1} \quad .
$$
At the same time the channel $P \rightarrow \overline{\nu}K^+$ is strongly
suppressed in conventional $SU(5)$~\cite{su5}~.
 
Both predictions of the $SU(5)$ GUT are now known to be inconsistent with experiment. The gauge coupling constants, as measured in LEP, do not unify in terms of conventional $SU(5)$ and the rate of   $P  \to e^+\pi^0$ is above the experimental upper limits~\cite{pdg}  
$$
\Gamma_{exp} (P \to \epi)  < \frac{1}{9}\  10^{-32 }  \quad yrs^{-1} \quad .
$$  
Making  $SU(5)$ supersymmetric \ \cite{ss5}, which leads to unification of the LEP gauge
coupling constants at the scale of $\approx 10^{16}\, GeV$\cite{uni}, can solve
these problems for an effective SUSY breaking below $1\
TeV$\footnote{SUSY can also stabilize the hierarchy, obviously.}. The
Yukawa unification \qquad
$m_\tau (GUT) \simeq m_b(GUT)$ \qquad also works well in SUSY $SU(5)$\ . The
high unification scale suppresses strongly the gauge boson mediated
$D=6$ proton decay in SUSY GUTs and avoids the above conflict with
experiment. On the other hand, SUSY $SU(5)$ allows for $D=5$    
contributions to the proton decay via a color triplet Higgs exchange. In terms of the $SU_C(3)$ indices this $D=5$ effective coupling reads 
\be
\epsilon_{ijk} q_i q_j \tilde{q}_k \tilde{\ell}  \qquad .
\ee
The antisymmetry requires the contribution of at least two families
and because the c-quark  is too heavy the amplitude always involves an
s-quark. Hence, $D=5$ induced proton decay in SUSY theories is
dominated by the Kaon--channels. The rate of the $D=5$ contribution is
in general much too high. It can however be reduced in certain models~\cite{pd1} using small mixing angles and coupling constants.
One goes, in certain review papers, as far as saying that observations of Kaon--dominant proton decay will be a decisive evidence for SUSY. 
 
In the meantime it was shown~\cite{int}, that also in L-R symmetric GUTs like
$SO(10)$\ ~\cite{so10} or $E_6$\ ~\cite{e6}, the use of an intermediate breaking scale around
$10^{11}\, GeV\ $\,\footnote{Note, that this scale is not arbitrary,
  it is relevant also for the
  axion window, the Baryon asymmetry and the RH neutrinos.} leads to
gauge and Yukawa unification in a way similar to the SUSY GUTs. Also, as the scale of unification is higher, there is no problem here with the rate of  $P  \to e^+\pi^0$.
 
The aim of this paper is to show that if one makes  the L-R symmetric
GUTs $P_{LR}$ invariant, the corresponding non-SUSY GUTs also give
proton decay results similar to the SUSY ones. In particular, they
predict dominance of the proton decay K--channels with rates in the
range of observability by superKamiokande and ICARUS. The reason for this is that $P_{LR}$ invariance in the two light families will be shown to be  equivalent to a full right-handed (RH) mixing of these families and this practically leads to the exchange $d^c  \leftrightarrow s^c$ in the proton decay amplitude. 
It is important to note here that, as long as the predictability of
the proton decay is considered, pure GUT models are more reliable than
SUSY GUTs. The proton decay in non-SUSY GUTs is induced by gauge
bosons with known masses and coupling. In SUSY-GUT models for proton
decay there are, in contrast, many unknown parameters. In particular,
knowledge about the physics above the GUT scale is required \ ~\cite{pd1}. 

What is $P_{LR}$ ? \\
It is a local version of parity  invariance. 
$P_{LR}$ invariance under the product gauge group  $G_L \times G_R \times P_{LR}$ means invariance under parity accompanied simultaneously by the exchange of the groups $G_L \longleftrightarrow G_R$ , or explicitly
\be
P_{LR}\  (i,k)_{LH}\  {P_{LR}}^{-1}  = (k,i)_{RH} \quad .
\ee
This is analogous to the ``old'' assignment of baryons to the P
invariant representation $(\bar3,3) \oplus (3,\bar3)$ under the
global chiral  $SU_L(3)\times SU_R(3)$~\cite{GM}. The baryons then acquire their masses when $(\bar3,3) \oplus (3,\bar3)$  is broken down to the diagonal group $SU(3)_{L+R}$. Under this diagonal group they constitute ${\bf 8} + {\bf 1}$ Dirac spinors.  
However, while P
invariance for the global symmetry leads per definition to 
$ SU_L(3) \leftrightarrow SU_R(3)$ exchange, in the local case
$G_{L,R}$ is only an historical notation. The chirality of the local
currents is fixed by the representation content of the fermions under
$G_{L,R}$. Therefore, the exchange $G_L \leftrightarrow G_R$ is an
additional requirement for local gauge theories which completes the
analogy with the global case.

$P_{LR}$ invariant local gauge theories were constructed first for leptons\ci{pl} and then for all fermions in terms of $SU_C(3) \times SU_L(3)\times SU_R(3)$ or $E_6$ \ci{pe6}.\\ 
It can also be applied to $SO(10)$~\ci{key} in terms of a L-R symmetric subgroup like the Pati-Salam one.\\ 
The ${\bf 16}$ fermionic family of $SO(10)$ GUT transforms, in this
case, under
$$
SO(10) \supset SU_C(4) \times SU_L(2)\times SU_R(2)\ ,
$$
as follows,
$$
{\bf 16}_{LH} = (4,2,1)_{LH}  \oplus  (\bar4,1,2)_{LH} \equiv{\bf
  f}_{LH} + \hat{{\bf f}}_{LH} 
$$
and 
$$ 
{ \overline{\bf 16}}_{RH} = (\bar4,2,1)_{RH}  \oplus  (4,1,2)_{RH} \quad .
$$
Applying  $P_{LR}$  
 
\be
(4,2,1)_{LH}  \longleftrightarrow (4,1,2)_{RH}  \qquad 
(\bar4,1,2)_{LH}  \longleftrightarrow (\bar4,2,1)_{RH} \quad .
\ee
 
Hence, to have an $SO(10) \otimes P_{LR}$ irreducible representation one
 needs two
{\bf 16}'s. In fact, using ${\bf 16}_{LH} \sim \overline{\bf 16}_{RH}$, 
 
\be
\begin{array}{lcll}
\Psi^1_{16LH}  \oplus \Psi^2_{16LH} &\sim & {\bf 16}^1_{LH} \oplus {\ov{\bf 16}}^2_{RH} & =  (4,2,1)^1_{LH} +  (\bar4,1,2)^1_{LH} \\[5pt]
  &   &   &  +\, (4,1,2)^2_{RH} + (\bar4,2,1)^2_{RH} \qquad .
\end{array}
\ee
 
Taking the families,\  $i,j = 1,2$,\  into account, we then have 
 
\begin{equation} 
P_{LR}\ {\bf f}^i(x)\  P^{-1}_{LR} = \epsilon^{ij} \sigma_2\  \hat{{\bf f}}^{j\star} (\bar x) \quad .
\end{equation}
 
The $P_{LR}$ invariant Yukawa coupling  mixes, therefore, the two $SO(10)$ representations\footnote{Note, that this form of the Yukawa Lagrangian can be fixed also by a discrete symmetry, independent of $P_{LR}$ \ .}
 
\begin{equation}
{\cal L}_Y = y_{12}  {\ov \Psi}^{1c} \Phi_{12} \Psi^2  + 
                      y_{21}  {\ov\Psi}^{2c} \Phi_{21} \Psi^1  +  h.c. \label{yuk}
\end{equation}
 
The relevant Higgs representations  $\Phi_{12}, \Phi_{21} $  transform under $SO(10)$ like 
$$
{\bf 16}  \otimes {\bf 16} = ({\bf 10} \oplus {\bf 126})_S \oplus {\bf 120}_{AS} \qquad .
$$
In terms of $SU_C(4) \times SU_L(2)\times SU_R(2)$ we need $\Phi \sim (1,2,2)$ to have 
 
$$
{\ov\Psi}^c \Phi \Psi \sim (\bar4,2,1) \otimes (1,2,2) \otimes (4,2,1) + \ldots
$$
 
Such contributions are included only in the ${\bf 10}_S$ and ${\bf
  120}_{AS}$  but not in ${\bf 126}_S$ \, \footnote{which is not present
  in superstring inspired $SO(10)$\ci{ssi}.}. Both symmetric and antisymmetric representations must be used to obtain non degenerate masses. 
 
To make the model realistic, we have to introduce the third heavy family, 
together with the known LH mixing including CP violation and to explain the smallness of the neutrino masses. The last problem can be solved, as usual, by giving the RH neutrinos heavy Majorana masses when $SO(10)$ will be broken down to the intermediate Pati-Salam gauge group. 
 
Now, the third family is not directly involved in the proton
decay. Its LH mixing with the light families is very small and CP
violation will have a little effect only on the proton decay. So let
us first discuss the two light families only and start with an exact
$P_{LR}$  symmetry.

The mass matrices must look as follows:

$$
\begin{array}{cccc}
\left (
\begin{array}{cc}
0&m_u\\
m_c&0
\end{array} \right ) &
\left (
 \begin{array}{cc}
0&m_d\\
m_s&0
\end{array} \right ) &
\left (
\begin{array}{cc}0&m_e\\
m_\mu&0
\end{array} \right ) &
\left (
\begin{array}{cc}
0&m_{\nu_e}\\
m_{\nu_\mu}&0
\end{array} \right ) \qquad .
\end{array}
$$
 
The diagonalization of these matrices obviously gives us a full RH mixing and vanishing LH mixing angles so that the physical families are then constructed as follows
 
\begin{equation}
{\bf f}^{e} = (4,2,1)^{1}_{LH} \oplus (4,1,2)^{2}_{RH} \sim (4,2,1)^1_{LH} \oplus (\bar 4,1,2)^2_{LH}
 \end{equation}

\begin{equation}
{\bf f}^{\mu} = (4,2,1)^{2}_{LH} \oplus (4,1,2)^{1}_{RH} \sim (4,2,1)^2_{LH} \oplus (\bar 4,1,2)^1_{LH}
 \end{equation}
 
Practically speaking, the RH components of the initial representations relevant for the proton decay are exchanged:
 
\begin{equation}
u^c_{LH} \llra c^c_{LH}  \qquad \qquad  d^c_{LH} \llra s^c_{LH}  \qquad \qquad  e^+_{LH} \llra \mu^+_{LH} \label{excg} 
\end{equation}
 
The replacement of $\hat{\nu}_e  \leftrightarrow   \hat{\nu}_\mu$ does not play a role for the proton decay as the RH neutrinos must be very heavy to induce the see-saw mechanism.

The RH mixing angles are not observable in the Standard Model
(SM). This means that the pure diagonal and pure off-diagonal matrices
are equivalent on the SM level. This is clearly not true on the GUT scale. In
particular, the RH mixing plays an important role for the \pd. Another
GUT effect for which the RH mixing plays a major role is the Baryon
asymmetry. Especially if it is included by leptogenesis \ci{lg},
i.e. via decay of the heavy RH neutrinos. 
 
The predictions of \plr\  invariance for the \pd\  are obtained
 making the replacement (\ref{excg}) in the conventional $SO(10)$ effective
 Lagrangian \ci{su5}. Neglecting terms involving heavy fermions we then obtain

\begin{eqnarray}
{\cal{L}}_{eff}& = &- \frac{4G}{\sqrt{2}} \epsilon_{\alpha\beta\gamma}
[\ov{u^{c\gamma}_L} \gamma^\lambda s^\beta_L (\ov{\nu^c_{\mu R}}\gamma_\lambda
d^\alpha_{R} +  \ov{\mu^+_L}  \gamma_\lambda  u^\alpha_L)] \nonumber \\ [10pt] &
& +
\frac{4G'}{\sqrt{2}} \epsilon_{\alpha\beta\gamma} [(\ov{s^{c\gamma}_L}
\gamma^\lambda  d^\beta_L + \ov{d^{c\gamma}_L}  \gamma^\lambda s^\beta_L)
\ov{\nu^c_{\mu R}} \gamma_\lambda   u^\alpha_R 
-  \ov{s^{c\gamma}_L}
 \gamma^\lambda u^\beta_L ( \ov{\mu^+_R}\gamma_\lambda
u^\alpha_R)] \label{leff}\quad ,
\end{eqnarray}
 
where $G = \frac{g^2}{2M^2_{x,y}}$   and  $G' = \frac{g^2}{2M^2_{x',y'}}$     are  the effective ``Fermi coupling constants'' for the $SO(10)$ leptoquark gauge bosons $X,Y$  and  $X',Y'$. 
 
Clearly, only the following \pd \  modes 
$$
P \lra  \nmk  \qquad \hbox{and}  \qquad  P \lra \muk
$$
are possible. For the B-violating neutron decay eq. (\ref{leff}) gives only
$$
N \lra \bar\nu_\mu K^0 \qquad \hbox{and} \qquad 
N  \lra  \bar\nu_
\mu K^{\star 0} .
$$
 
We have not considered until now the observed family mixing, i.e. the
Cabibbo angle. To take this into  account we must break \plr\
invariance in the Yukawa Lagrangian (\ref{yuk}). In this
paper, however, we will discuss only the simplest possible
perturbation of the above off-diagonal matrices that gives the LH
Cabibbo mixing. This means adding a (2,2) element in the $M_d$  or
the $M_u$ matrix such that the other matrices remain off-diagonal: 
 
$$
\begin{array}{ccc}
\hbox{\hphantom{I}I)} \qquad M_u = \left (
\begin{array}{cc}
0&m_u\\
m_c&0
\end{array}   \right )& 
  \qquad M_d =   \left (
\begin{array}{cc}
0&a\\
b&c
\end{array} \right ) &
 \qquad M_e=   \left (
 \begin{array}{cc}
0&m_e\\
m_\mu&0
\end{array} \right ) 
\end{array} \\
$$
$$
\begin{array}{ccc}
\hbox{II)} \qquad M_u =   \left (
\begin{array}{cc}
0&\alpha\\
\beta&\gamma
\end{array}   \right )& 
  \qquad M_d =   \left (
\begin{array}{cc}
0&m_d\\
m_s&0
\end{array} \right ) &
 \qquad M_e=   \left (
 \begin{array}{cc}
0&m_e\\
m_\mu&0
\end{array} \right ) 
\end{array}
$$

We will take the matrix elements to be real, in view of the previous remark on CP violation. 
 
The requirement that the LH mixing will be the Cabibbo one then fixes also the RH mixing. E.g. for the above cases we find
 
$$
\hbox{\hphantom{I}I)} \qquad \tan \Theta^d_R  =  - \frac{m_d (\mu = 1 GeV)}{m_s(\mu = 1 GeV)}  \cot  \Theta_{Cabibbo}    \simeq -0.228
$$
$$
\hbox{II)} \qquad \tan \Theta^u_R  =  \hphantom{-} \frac{m_u (\mu = 1 GeV)}{m_c(\mu = 1 GeV)}  \cot  \Theta_{Cabibbo}    \simeq  0.017
$$
 
Note, that the $SO(10)$  model dictates the mass matrices at the GUT scale.
Those masses must therefore be renormalized down to low energies.  
The mixing angles are, however, quite scale independent. This is related  
to the fact that the renormalization gives practically the same factor
for the whole matrix. We shall use the masses at the scale  $\mu = 1\  GeV$ as it is usually done.
 
As we look for small deviations from the \plr \  invariant case, let us
introduce the most general mixing of the quarks in the nucleon decay
Lagrangian eq. (\ref{leff}).

\renewcommand{\arraystretch}{1.5}
\beq
u'_{L,R} & = & \cos(\alpha_{L,R})u_{L,R} + \sin(\alpha_{L,R})c_{L,R} \\ 
c'_{L,R} & = &- \sin(\alpha_{L,R})u_{L,R} + \cos(\alpha_{L,R})c_{L,R} \\
d'_{L,R} & = & \cos(\beta_{L,R})d_{L,R} + \sin(\beta_{L,R})s_{L,R} \\
s'_{L,R} & = & -\sin(\beta_{L,R})d_{L,R} + \cos(\beta_{L,R})s_{L,R}
\eeq

Using this and the  Fierz-identities we can write ${\cal{L}}_{eff}$ in
the form
\clearpage
\beq
{\cal L} & = & A_1\bar u^c \gamma_L u \bar e^+ \gamma_Ld + 
        A_2\bar u^c \gamma_L u \bar e^+ \gamma_Rd \nonumber \\
 & + & A_3\bar u^c \gamma_L u \bar \mu^+ \gamma_Ld + 
        A_4\bar u^c \gamma_L u \bar \mu^+ \gamma_Rd  \nonumber\\
 & + & A_5\bar u^c \gamma_L u \bar e^+ \gamma_Ls + 
        A_6\bar u^c \gamma_L u \bar e^+ \gamma_Rs \nonumber \\
 & + & A_7\bar u^c \gamma_L u \bar \mu^+ \gamma_Ls + 
        A_8\bar u^c \gamma_L u \bar \mu^+ \gamma_Rs \nonumber \\
 & + & A_9\bar u^c \gamma_L d \bar \nu^c_e \gamma_Rd + 
       A_{10}\bar u^c \gamma_L d \bar \nu^c_\mu \gamma_Rd \nonumber \\
 & + & A_{11}\bar u^c \gamma_L d \bar \nu^c_e \gamma_Rs + 
       A_{12}\bar u^c \gamma_L d \bar \nu^c_\mu \gamma_Rs \nonumber \\
 & + & A_{13}\bar u^c \gamma_L s \bar \nu^c_\mu \gamma_Rd + 
       A_{14}\bar u^c \gamma_L s \bar \nu^c_e \gamma_Rd + h.c. 
\eeq
  
where $\gamma_{L,R}=\gamma_\mu (1 \mp \gamma_5)$
and the $A_i$ parameters are defined as follows:

\beq
A_1 &=& G\sin (\alpha_R + \alpha_L) \sin(\beta_L) - 
        G\sin(\alpha_R + \beta_L) \sin(\alpha_L) \nonumber \\
A_2 &=& G\sin (\alpha_R + \alpha_L) \sin(\beta_R) - 
        G'\sin(\alpha_L + \beta_R) \sin(\alpha_R) \nonumber \\
A_3 &=&- G\sin (\alpha_R + \alpha_L) \cos(\beta_L) + 
        G\sin(\alpha_R + \beta_L) \cos(\alpha_L) \nonumber \\
A_4 &=&- G\sin (\alpha_R + \alpha_L) \cos(\beta_R) + 
        G'\sin(\alpha_L + \beta_R) \cos(\alpha_R) \nonumber \\
A_5 &=&- G\sin (\alpha_R + \alpha_L) \cos(\beta_L) + 
        G\cos(\alpha_R + \beta_L) \sin(\alpha_L) \nonumber \\
A_6 &=&- G\sin (\alpha_R + \alpha_L) \cos(\beta_R) + 
        G'\cos(\alpha_L + \beta_R) \sin(\alpha_R) \nonumber \\
A_7 &=&- G\sin (\alpha_R + \alpha_L) \sin(\beta_L) - 
        G\cos(\alpha_R + \beta_L) \cos(\alpha_L) \nonumber \\
A_8 &=&- G\sin (\alpha_R + \alpha_L) \sin(\beta_R)- 
        G'\cos(\alpha_L + \beta_R) \cos(\alpha_R) \nonumber \\
A_9 &=&- G\sin (\alpha_R + \beta_L) \sin(\beta_R) + 
        G'\sin(\beta_L + \beta_R) \sin(\alpha_R) \nonumber \\
A_{10} &=& G\sin (\alpha_R + \beta_L) \cos(\beta_R) - 
        G'\sin(\beta_L + \beta_R) \cos(\alpha_R) \nonumber \\
A_{11} &=& G\sin (\alpha_R + \beta_L) \cos(\beta_R) - 
        G'\cos(\beta_L + \beta_R) \sin(\alpha_R) \nonumber \\
A_{12} &=& G\sin (\alpha_R + \beta_L) \sin(\beta_R) + 
        G'\cos(\beta_L + \beta_R) \cos(\alpha_R) \nonumber \\
A_{13} &=&- G\cos (\alpha_R + \beta_L) \cos(\beta_R) + 
        G'\cos(\beta_L + \beta_R) \cos(\alpha_R) \nonumber \\
A_{14} &=& G\cos (\alpha_R + \beta_L) \sin(\beta_R) - 
        G'\cos(\beta_L + \beta_R) \sin(\alpha_R)  
\eeq

Note, that here $\alpha_R, \beta_R$ are the deviations from the full RH mixing, i.e. 
$$
\alpha_R = \frac{\pi}{2} - \Theta^u_R  \qquad ,\qquad  
\beta_R = \frac{\pi}{2} - \Theta^d_R  \qquad .
$$
We will use the calculations done in\ci{key}, \ci{sus} in terms of the $SU(6)$ quark model of Gavela et al\ci{gavela}. I.e. the partial rates of the different channels are given as follows

\be
\Gamma = \left[ \frac{M^5_p}{(10^{14})^4}\right] \left(
  \frac{10^{14}}{M_X}\right)^4 \rho\ 16\pi \alpha^2_U \left[ 
  \frac{|A_3|^2|\psi(0)|^2}{M^3_p}\right]
 \left [|A_i^L|^2\langle|M_L|^2\rangle|A^L_{12}|^2 + |A_i^R|^2
\langle|M_R|^2\rangle|A^R_{12}|^2\right]
\ee

This expression depends on the $(X,Y),(X', Y')$  masses, the unification scale, $M_U$ and the unification coupling constant of $SO(10), \ \alpha_U$\ .
$M_U \   \hbox{and}\  \alpha_U$  were already calculated for
$SO(10)$ broken down to the intermediate gauge group $SU_C (4) \otimes
SU_L(2) \otimes SU_R(2)$ in several papers \ci{int}. 
As for $(X,Y)\quad \hbox{and}\quad (X', Y')$, their masses are
expected to be around $M_U$. Threshold effects may, however, lead to
somewhat smaller $M_{x,y}\  \hbox{or\ } M_{x\prime,y\prime}$\ . To
take this into account and to get an idea how much such threshold
corrections effect the partial rates, we studied three possibilities: 

$$
\renewcommand{\arraystretch}{1.5}
\begin{array}{cccccc}
(i) & M^2_{x,y} &=&    M^2_{x\prime,y\prime} &=& M^2_U \\
(ii) & M^2_{x,y} &= &  10  M^2_{x\prime,y\prime} &= &M^2_U \\
(iii) &  M^2_{x\prime,y\prime} &= &10  M^2_{x,y} &= & M^2_U
\end{array}
$$
 
For ${\left |\Psi(0) \right|}^2$ we took\ci{kroll}: 
$$
{\left |\Psi(0) \right|}^2 = 0.012\ GeV^2 \qquad .
$$

The phase space factor $\rho$ was estimated \`a la Karl and Kane\ci{kk} as 
$$
\rho = (1-\chi^2)(1-\chi^4)
$$
where
$$
\chi = \frac{m_{meson}}{m_{nucleon}} \qquad .
$$ 
For $\left | A^{L,R}_{12} \right |$ we used \ci{je} 

\be
|A^L_{12}|^2 = \left[ \left( \frac{\alpha_2 (M^2_Z)}{\alpha_U}\right)^
{\frac{27}{(86-8N_f)}}\quad 
\left( \frac{\alpha_1 (M^2_Z)}{\alpha_U}\right)^
{\frac{-69}{(6+40N_f)}}\right]^2 \nonumber 
\ee
\be
|A^R_{12}|^2 = \left[ \left( \frac{\alpha_2 (M^2_Z)}{\alpha_U}\right)^
{\frac{27}{(86-8N_f)}}\quad 
\left( \frac{\alpha_1 (M^2_Z)}{\alpha_U}\right)^
{\frac{-33}{(6+40N_f)}}\right]^2 \nonumber 
\ee
 
The renormalization factor $A_3$ was calculated using 

\be
A_3 = \left( \frac{\alpha_s(\mu^2)}{\alpha_U}\right)^{6/(33-2N_f)}
\ee
where
$$
\alpha_s(\mu^2) = (12\pi) / (25 \ln \frac{\mu^2}{\Lambda^2_{QCD}})
$$

with $\Lambda_{QCD} = 324{+20 \atop -56}\, MeV$\, \ci{wr}. 

As was explained before we considered two possibilities to introduce
the Cabibbo mixing into the \plr \ invariant formalism:

Case { }I) a (2,2) entry in $M_d$ \qquad\\ and \\
Case II)   a (2,2) entry in $M_u$ .

The calculated branching ratios (BRs) for the proton decay can be
found in tables \ref{tab1}, \ref{tab2} and the ones for the neutron decay in tables
\ref{tab3}, \ref{tab4}. In addition to the BRs in \% we give the total nucleon decay rate
in the different cases.

\begin{table}
\begin{center}
\begin{tabular}{|c|c|c|c|}
\hline
\hbox{channel} & $G' = \frac{1}{10}G$ & $G'=G$ & $G'=10G$ \\
\hline \hline
$\epi$ & 0 & 0 & 0  \\
$\et$ & 0 & 0 & 0  \\
$\ero$ & 0 & 0 & 0 \\
$\eo$ & 0 & 0 & 0  \\
$\nep$ & 5.7 & 2.1 & 0.03 \\
$\nero$ & 0.3 & 0.1 & 0.0 \\
$\muk$ & 43.2 & 16.6 & 1.3 \\
$\nmk$ & 3.9 & 1.4 & 0.02 \\
$\ek$ & 0 & 0 & 0 \\
$\mpi$ & 1.9 & 11.0 & 15.3 \\
$\mt$ & 0.4 & 2.2 & 3.0 \\
$\eks$ & 0 & 0 & 0 \\
$\mro$ & 0.1 & 0.6 & 0.9 \\
$\mo$ & 0.9 & 5.3 & 7.4 \\
$\nek$ & 36.7 & 13.6 & 0.2 \\
$\neks$ & 5.7 & 2.1 & 0.03 \\
$\nmks$ & 0.02 & 2.4 & 3.9 \\
$\nmro$ & 0.8 & 0.3 & 0.0 \\
$\nmpi$ & 0.3 & 42.2 & 67.9 \\
\hline \hline
\hbox{total rate}\ $(yrs^{-1})$ & $3.8\ 10^{-34}$ &$1.0\ 10^{-35}$  & $6.9\ 10^{-34}$   \\
\hline
\end{tabular}
\end{center}
\caption{Branching ratios and the total rate for proton decay channels
  in case (I) for the three different versions .\label{tab1}}
\end{table}

\begin{table}
\begin{center}
\begin{tabular}{|c|c|c|c|}
\hline
\hbox{channel} & $G' = \frac{1}{10}G$ & $G'=G$ & $G'=10G$ \\
\hline \hline
$\epi$ & 1.4 & 1.9 & 1.3  \\
$\et$ & 0.3 & 0.4 & 0.3  \\
$\ero$ & 0.08 & 0.1 & 0.07 \\
$\eo$ & 0.7 & 0.9 & 0.6  \\
$\nep$ & 0 & 0 & 0 \\
$\nero$ & 0 & 0 & 0 \\
$\muk$& 0.0 & 0.0 & 0.0 \\
$\nmk$ & 0.0 & 0.0 & 0.0 \\
$\ek$ & 46.3 & 41.4 & 28.2 \\
$\mpi$ & 6.4 & 6.8  & 0.2  \\
$\mt$ & 1.3 & 1.3 & 0.04\\
$\eks$ & 0.4 & 0.4 & 0.2\\
$\mro$ & 0.4 & 0.4 & 0.01 \\
$\mo$ & 3.1 & 3.3 & 0.1 \\
$\nek$ & 8.5 & 9.7 & 66.9 \\
$\neks$ & 0.8 & 1.6 & 0.9 \\
$\nmks$ & 1.6 & 1.7 & 0.06 \\
$\nmro$ & 0.0  & 0.0  & 0.0 \\
$\nmpi$ & 28.7 & 30.2 & 1.1  \\
\hline \hline
\hbox{total rate}\ $(yrs^{-1})$ & $1.7\ 10^{-33}$ & $1.6\ 10^{-35}$ & $4.4\ 10^{-34}$ \\
\hline
\end{tabular}
\end{center} 
\caption{Branching ratios and the total rate for proton decay channels
 in case (II) for the three different versions .\label{tab2}}
\end{table}

\begin{table}
\begin{center}
\begin{tabular}{|c|c|c|c|}
\hline
\hbox{channel} & $G' = \frac{1}{10}G$ & $G'=G$ & $G'=10G$ \\
\hline \hline
$\epin$ & 0 & 0 & 0 \\
$\mpin$ & 1.2 & 10.6 & 20.5 \\
$\nepn$ & 0.9 & 0.5 & 0.01 \\
$\netn$ & 0.2  & 0.1 & 0.0 \\
$\nekn$ & 26.6 & 14.6 & 0.3 \\
$\nmpin$ & 0.06 & 10.2 & 22.7 \\
$\nmtn$ & 0.01 & 2.6 & 5.8 \\
$\nmkn$ & 5.9 & 3.2 & 0.07 \\
$\eron$ & 0 & 0 & 0 \\
$\mron$ & 0.08 & 0.7 & 1.4 \\
$\neron$ & 0.2 & 0.1 & 0.0 \\
$\neon$ & 1.7 & 0.9 & 0.02 \\
$\neksn$ & 57.6 & 31.7 & 0.6 \\
$\nmron$ & 0.01 & 2.1 & 4.6 \\
$\nmon$  & 0.1 & 18.4 & 41.1 \\
$\nmksn$ & 5.3 & 2.9 & 0.06 \\
\hline \hline
\hbox{total rate}\ $(yrs^{-1})$ & $1.2\ 10^{-33}$ & $2.1\ 10^{-35}$ & $1.0\ 10^{-33}$ \\
\hline
\end{tabular}
\end{center}
\caption{Branching ratios and the total rate for neutron decay
  channels in case (I) for the three different versions .\label{tab3}}
\end{table}

\begin{table}
\begin{center}
\begin{tabular}{|c|c|c|c|}
\hline
\hbox{channel} & $G' = \frac{1}{10}G$ & $G'=G$ & $G'=10G$ \\
\hline \hline
$\epin$ & 3.0 & 2.6 & 2.7 \\
$\mpin$ & 13.4 & 9.4 & 0.5 \\
$\nepn$ & 0 & 0 & 0 \\
$\netn$ & 0  & 0 & 0 \\
$\nekn$ & 23.8 & 16.9 & 0.9 \\
$\nmpin$ & 14.9 & 10.6 & 0.6 \\
$\nmtn$ & 3.8 & 2.7 & 0.1 \\
$\nmkn$ & 0.0 & 0.0 & 0.0 \\
$\eron$ & 0.6 & 0.5 & 0.6 \\
$\mron$ & 1.3 & 1.0 & 0.4 \\
$\neron$ & 0 & 0 & 0 \\
$\neon$ & 0 & 0 & 0 \\
$\neksn$ & 7.7 & 34.1 & 93.2 \\
$\nmron$ & 3.0 & 2.1 & 0.1 \\
$\nmon$  & 27.0 & 19.1 & 1.1 \\
$\nmksn$ & 0.01 & 0.0 & 0.02 \\
\hline \hline
\hbox{total rate}\ $(yrs^{-1})$ & $1.6\ 10^{-33}$ & $2.3\ 10^{-35}$ & $4.1\ 10^{-34}$ \\
\hline
\end{tabular}
\end{center}
\caption{Branching ratios and the total rate for neutron decay
  channels in case (II) for the three different versions .\label{tab4}}
\end{table}

Considering the tables we see, as was expected, that in all versions
(except for the Iii one) the Kaon decay channels dominate. This is
especially true for the versions Iiii and IIii.\\
All three versions of case (I) have a considerable rate for the channel 
\footnote{It is interesting to note that this mode is
     also enhanced in the SUSY GUT model of ref.\ci{s3}. In this SUSY
model the two light families are also symmetric, they transform as $\bf2$
under the flavor $[S_3]^3$ symmetry.} 
$$
     P\ra \ek 
$$    
The versions Ii and Iii also have a considerable contribution to the
$$
      P\ra \nmpi
$$
channel. The interesting ''new" neutron decay mode is
$$
      N\ra \nmon
$$ 
which is present in most versions.

The rates we calculated used the central values of the parameters,
without taking into account all kind of corrections and in particular
not the full threshold effects. Those corrections were estimated
already in a paper of Lee et al\ci{lee} for the standard $SO(10)$
broken down to Pati-Salam gauge group. They found  
 
$$
\tau_{P \to  \epi} =  1.44  \times 10^{37.4 \pm 0.7 \pm 1.0   {+0.5
    \atop -5.0}} yrs.
$$
Or explicitly,

$ \pm 0.7 $ \quad - comes from the proton decay matrix element
evaluation \ci{su5},

$\pm 1.0$ \quad - comes from uncertainty in the LEP data, while.

${+0.5 \atop -5.0}$ \quad  - in due to the threshold corrections.
 
This gives an idea of the possible corrections to the absolute rates
given above. The branching ratios are, however, independent of those
corrections. Taking those into account, it is clear that all the
interesting decay modes fall into the range of observability of
superKamiokande and ICARUS, i.e. $ O(10^{34})$ yrs.\\
The versions with $G'\not= G$ have, however, essentially larger
absolute rates\ ( by a factor $\sim 100$ ) and have a better chance to
be tested experimentally. 

The minimal models we considered here are not complete in the sense
that the heavy family is not explicitly included. To introduce the
third family while keeping the approximate \plr\  invariance for the two light families, there are two directions one can take:
 
a) To keep the special role of the top-family in getting its masses
directly from the Higgs  mechanism.\\ Models of this kind assume that the three families transform as  ${\bf 2}\oplus {\bf 1}$ under a
family group like $O(3),\  U(2) \  \hbox{or} \  (S_3)^3$\ci{u2},\ci{s3}
. The
``perturbations'', like radiative corrections or non-renormalizable
terms, which give masses to the light families in those models must then be, in our case, (approximately) \plr \  invariant.  
 
b) To use a very heavy fourth family which obeys also \plr\
invariance with the third one and has very small mixing angles with
the light families.\\In this way one can obtain predictions for the
masses of the very heavy fermions.  
 
Those possibilities will be studied in detail in another paper \ci{am}. 
Let us only mention here that in the case a) we have to start from 
$3 \times 3$  mass matrices of the form

\be
\left (
\begin{array}{ccc}
0&a&0 \\
b&0&0 \\
0&0&d       
\end{array} \right ) \qquad   .
\ee

It is natural then  to look for nearest neighbour interactions
(sometimes called non-hermitian Fritzsch matrices \ci{frit}) which
give the right fermionic masses and Cabibbo-Kobayashi-Maskawa  LH
mixing angles. In the framework of the SM, a given mass matrix can
always be transformed into a next neighbour interaction matrix
\ci{bran}, but on the GUT scale this will define the RH mixing angles
that are relevant for nucleon decay.       

%%referenzen%%

\bye
\end{document}